\documentclass{sig-alternate}
\usepackage{mathptmx} 

\usepackage{fancyhdr}
\usepackage[normalem]{ulem}
\usepackage[hyphens]{url}
\usepackage[sort,nocompress]{cite}
\usepackage[final]{microtype}
\usepackage{flushend}
\newcommand{\ignore}[1]{}
\usepackage{fancyhdr}
\usepackage[normalem]{ulem}
\usepackage[hyphens]{url}
\usepackage{microtype}

\usepackage[sort,nocompress]{cite}
\usepackage{flushend}
\usepackage{listing}
\usepackage{fancyvrb}
\usepackage{units}
\usepackage{makecell}
\usepackage{paralist}
\usepackage{graphicx}
\usepackage{comment}
\usepackage{color}
\usepackage{todonotes}
\usepackage{xspace}
\usepackage{enumitem}
\usepackage{wrapfig}

\usepackage[bookmarks=true,breaklinks=true,letterpaper=true,colorlinks,linkcolor=black,citecolor=blue,urlcolor=black]{hyperref}

\fancypagestyle{firstpage}{
  \fancyhf{}
  
  \fancyfoot[C]{\thepage}
}

\newcommand\punt[1]{}

\newlength{\figurenegativevspace}
\newcommand{\name}{\textsf{MI6}\xspace}
\newtheorem{property}{Property}

\title{MI6: Secure Enclaves in a\\ Speculative Out-of-Order Processor}
\author{Thomas Bourgeat\thanks{Student authors listed in alphabetical order.}\and Ilia Lebedev$^*$ \and Andrew Wright$^*$ \and Sizhuo Zhang$^*$ \and {Arvind~~~~~~~~~~Srinivas Devadas}\\\mbox{}\\
MIT Computer Science and Artificial Intelligence Laboratory}

\begin{document}
\maketitle
\thispagestyle{firstpage}
\pagestyle{plain}

\begin{abstract}
    Recent attacks have broken process isolation by exploiting microarchitectural side channels that allow indirect access to shared microarchitectural state. Enclaves strengthen the process abstraction to restore isolation guarantees.

We propose \name, an aggressively speculative out-of-order
processor capable of providing secure enclaves under a threat model
that includes an untrusted OS and an attacker capable of
mounting any software attack currently considered practical, including
those utilizing control flow mis-speculation.
\name is inspired by Sanctum \cite{costan2016sanctum} and
extends its isolation guarantee to more realistic memory hierarchy.
It also introduces a \texttt{purge} instruction, which is used only when a secure process is (de)scheduled, and implements it for a complex processor microarchitecture.
We model the performance impact of enclaves in \name through FPGA emulation on
AWS F1 FPGAs by running SPEC CINT2006 benchmarks as enclaves within an untrusted Linux OS.
Security comes at the cost of approximately 16.4\% average slowdown for protected programs. 


\end{abstract}

\section{Introduction}
\label{sec:intro}

\subsection{Secure Enclaves}

The process abstraction is pervasive and underpins modern software systems.
Conventional wisdom pertaining to process security teaches that an attacker that cannot \emph{name} a particular state element cannot attack that element.
Even in a situation where unconditional trust in the OS is considered tenable, the architectural process isolation an OS can enforce with commodity hardware falls short: microarchitectural side channels allow indirect access to shared microarchitectural state from which information about a process can be leaked (e.g., \cite{bonneau2006cache}, \cite{yarom2013llctiming}).
Recent attacks based on control flow speculation (e.g., Spectre \cite{Kocher2018spectre}) have brought this issue to the fore.

Our focus in this paper is to strengthen the capability of conventional systems,
i.e., speculative out-of-order (OOO) multicores running an OS (e.g., Linux), with processes with strong
isolation guarantees.
We will call such processes \emph{enclaves}, borrowing the term from Intel SGX \cite{mckeen2013sgx}, though the idea predates SGX.
Enclaves are to be used to run secure tasks, and will coexist with ordinary
processes.


Strong isolation requires a stronger guarantee than private memory.
The goal of the enclave abstraction is to achieve the following property:
\begin{property}[Strong Isolation]
Any attack by a privileged attacker program co-located on a machine with the victim enclave that can extract a secret inside the victim, can also be mounted successfully by an attacker on a different machine than the victim.
\label{security-property}
\end{property}

The attacker on a different machine than the victim is only able to communicate
with the victim through the victim's public API and can observe the latency of these API calls.
No other program should be able to infer anything private from the enclave program through its use of shared resources or shared microarchitectural state, e.g., cache or branch predictor state.
This implies that it is not enough to just give a unique set of addresses to an enclave, but separation of resources has to be provided at all levels of the cache hierarchy where these addresses may reside.
Our goal is to make sure that an enclave cannot leak information to or be
influenced by any program running in the system. To achieve this,
all shared resources in our microarchitecture are isolated \textit{spatially} and
\textit{temporally} as required by our threat model, described in Section~\ref{subsec:attacker}.

Enclaves trade expressivity for security; enclave software is restricted in how it interacts with untrusted software, including the OS (cf. Section \ref{subsec:abstraction}).
A trusted \textit{security monitor} running in machine mode
mediates enclave entry and exit, and
verifies resource allocation decided by untrusted system software to ensure, for example, allocation of non-overlapping memory to different enclaves.
The security monitor (cf. Section~\ref{subsec:security-monitor}) is designed to protect itself from tampering (by appropriately configuring hardware protection mechanisms), even if the OS has been compromised.

We trust our hardware and assume that it is bug free. Our threat model
assumes a privileged software adversary and is detailed in
Section \ref{subsec:attacker}; we also describe what is outside
our threat model in Section \ref{subsec:outside-threat}.
We are primarily concerned with attacks that exploit shared hardware
resources to interact with the victim enclave via methods
outside its public API.

\name is based on the open-source out-of-order RiscyOO~\cite{riscyoo-2018} processor,
and provides secure enclaves under our threat model.
We model the performance impact of enclaves in \name through FPGA emulation on
AWS F1 FPGAs by
running benchmarks from SPEC CINT2006 on top of an untrusted Linux OS.


\subsection{Contributions and organization}

Like Sanctum \cite{costan2016sanctum}, we argue that enclaves defined as a strengthened process are an excellent abstraction for secure computation.  We make the following contributions in support of our argument:
\begin{compactenum}
\item We show that there are many subtle side channels associated with queues and associated arbitration required to handle multiple outstanding memory requests in the memory hierarchy, and describe how to enforce strong isolation in such systems (cf. Section \ref{sec:llc}). \name therefore protects against a broader class of attacks than Sanctum, which excludes, for example, attacks that use the cache coherence bandwidth and DRAM controller bandwidth channels from its threat model.

\item We show that the complexity of the out-of-order processor core can be completely decoupled from the complexity of a modern memory hierarchy by using a new \verb|purge| instruction, which can be easily incorporated in any Instruction Set Architecture (ISA). We describe optimizations based on indistinguishability to software in our \verb|purge| implementation (cf. Section \ref{subsec:purge-instruction}).

\item We describe key modifications to the security monitor to maintain strong isolation in a speculative processor (cf. Section \ref{subsec:security-monitor}), and an optimization relating to access permissions checking (cf. Section \ref{sec:steady:addr}).

\item We provide a detailed evaluation of the performance overhead of enclaves in an implementation of the \name OOO processor.
We show, for example, the cost of flushing shared microarchitectural state on each system call, which is required for strong isolation.
Overall, security comes at the cost of approximately 16.4\% average slowdown for protected programs. We note that this number assumes a baseline that is not protected against physical attacks on memory unlike Aegis \cite{suh2003aegis} and Intel SGX \cite{mckeen2013sgx}, and with a constant latency DRAM controller.

\end{compactenum}

\noindent {\bf Organization:}
In Section \ref{sec:enclave-threat}, we describe the enclave abstraction
and threat model.
Related work is the subject of Section \ref{sec:related}.
In Section \ref{sec:baseline} we describe the baseline implementation of the RiscyOO speculative out-of-order processor based on the RISC-V ISA \cite{waterman2014riscv} and list the hardware modifications we made.
Section \ref{sec:steady-state-isolation} describes how we provide steady state isolation of enclaves in \name.  Section \ref{sec:transition-protect-domains} describes how \name handles transitions between enclaves.
The performance of \name is evaluated in Section \ref{sec:evaluation}.
We conclude in Section \ref{sec:conc}.

\section{Enclaves and Isolated Execution}
\label{sec:enclave-threat}

In this section, we describe the enclave abstraction as presented in \cite{subramanyan2017formal} (cf. Section \ref{subsec:abstraction}).
We describe the high-level approach to (micro)architectural isolation \name employs to achieve Property~\ref{security-property} (cf. Section~\ref{subsec:isolation}).
We then describe the capabilities of the adversary (cf. Section~\ref{subsec:attacker}).
We also discuss what falls outside our threat model (cf. Section~\ref{subsec:outside-threat}).

\subsection{The Enclave Abstraction}
\label{subsec:abstraction}

A processor that can serve as an {\em enclave platform} implements isolated software environments in disjoint address spaces only accessible from within a given enclave's threads of execution.
An {\em enclaved process} resides entirely and exclusively in such isolated memory, which contains all code and data structures comprising the enclave, and is isolated from all other software in the system.

An enclave platform must guarantee integrity and private execution of an enclave in the presence of other software, as per its threat model, and thus strongly isolates the enclave from all other software. 
As a consequence of its isolation, an enclave cannot transparently receive system services or issue system calls, as software outside the enclave is not trusted with access to enclave private memory.
The enclave platform mediates control transfer to and from an enclave via statically-defined locations called entry points, and guarantees no side effects of execution remain across these context switches.

Enclaves requiring system services must proxy through untrusted software, and must respect that the OS services are untrusted.
With these guarantees, enclaves can become the only trusted components of an application (aside from the platform itself), and a carefully programmed enclave can execute safely even when privileged software is compromised.

The platform also implements the measurement and attestation protocol of \cite{secure-boot-2018} to prove enclave integrity to a remote party.


\subsection{Enclave Isolation}
\label{subsec:isolation}

The goal of the enclave abstraction is to achieve Property~\ref{security-property}.
To protect enclave integrity, the processor implements \emph{architectural} isolation (that of memory) by setting up invariants in a hardware mechanism to prevent all accesses to enclave-owned physical memory, allowing only the enclave's code access.
Any sharing of microarchitectural resources by mutually distrusting software may transmit private information via the availability of these finite resources, measurable via the timing of certain operations or other side channels.
Microarchitectural isolation is necessary to not only protect the confidentiality of enclave execution (for example closing side channels through cache state), but also to protect enclave integrity in the context of a core that executes speculatively.

Instead of directly implementing isolated enclaves, the \name hardware implements flushing, constraints on core instruction fetch, and a set of low-level isolation primitives (cf. Section \ref{sec:steady-state-isolation}) sufficient to partition each relevant sub-system of \name into ``protection domains'' that are non-overlapping allocations of machine resources.
When programs are running on multiple cores in different protection domains, in this steady state, \name guarantees non-interference and isolation of these domains.

The platform must also allow for transitions between different protection domain configurations in order to implement enclaves.
Following the example of~\cite{costan2016sanctum}, \name relies on a small trusted security monitor (cf. Section~\ref{subsec:security-monitor}), which executes in a dedicated protection domain to compose the machine's protection domains and constraints on execution into the high-level properties of isolated enclaves.
While the implementation details of an \name security monitor largely borrow from~\cite{sanctorum-2018}, and are out of scope in this manuscript, Section~\ref{subsec:security-monitor} details the aspects of the monitor relevant to the isolation of enclaves.


The trusted computing base (TCB) of the \name processor includes
the processor chip, memory (i.e., DRAM), as well as the security monitor binary.
The security monitor is the only software running in the \emph{machine mode}, which is the highest privilege mode in RISC-V and is more privileged than the \emph{supervisor mode} used by the untrusted OS \cite{waterman2015riscvsystem}.
\name isolates the software inside an enclave from other software on the same speculative out-of-order processor to satisfy Property \ref{security-property}.


\subsection{Attacker Capability}
\label{subsec:attacker}

We assume an insidious remote adversary able to exploit software vulnerabilities expected to be present in large system software.
Specifically, we assume that an attacker can compromise any operating system and hypervisor present on the computer executing the enclave, and can launch malicious enclaves.
The attacker has complete knowledge of the enclave platform's architecture and microarchitecture, and the software it loads.
The attacker can analyze passively observed data, such as page fault
addresses, as well as mount active attacks, such as memory probing, and cache tag state attacks (e.g., \cite{yarom2013llctiming}).
The attacker can exploit speculative state, branch predictor state,
and other shared microarchitectural state in any software attack, e.g.,
Spectre \cite{Kocher2018spectre}.


\subsection{What's Outside the Threat Model}
\label{subsec:outside-threat}

\name's isolation mechanisms exclusively address software attacks, and assume the absence of any adversary with physical access.
We do not protect against attacks such as \cite{brumley2005rsa}, where the victim application leaks information via its public API, and the leak occurs even if the victim runs on a dedicated machine.
We also consider any software attacks that rely on sensor data to be physical attacks.
\name does not protect against physical attacks on memory, but can be
augmented with memory encryption and integrity verification
exemplified by processors such as Aegis \cite{suh2003aegis} and
Oblivious Random Access Memory (ORAM) \cite{goldreich1987oram} \cite{stefanov2013path}
as in Ascend \cite{fletcher2012ascend} to variously defend against these attacks. ORAM overhead can be substantially reduced if smart memory is assumed \cite{invisimem,obfusmem}.
\name does not protect against denial-of-service (DoS) attacks, assumes correct underlying hardware, and does not protect against software attacks that exploit hardware bugs (fault-injection attacks such as rowhammer
\cite{kim2014rowhammer, google2015rowhammer}).
Finally, we exclude attacks that utilize shared performance counters.

\section{Related Work}
\label{sec:related}


\subsection{Microarchitectural side channel attacks}

Attacks that exploit microarchitectural side channels to leak information come in many varieties.
Attacks using cache tag state-based channels are known to retrieve cryptographic keys from a growing body of cryptographic implementations: AES\cite{osvik2006aes,bonneau2006aes}, RSA \cite{brumley2005rsa}, Diffie-Hellman \cite{kocher1996timing}, and elliptic-curve cryptography \cite{brumley2011ecc}, to name a few.
Such attacks can be mounted by unprivileged software sharing a computer with the victim software \cite{banescu2011cache}.

Sophisticated channel modulation schemes such as flush +reload \cite{yarom2014flush} and variants of prime+probe~\cite{liu2015llctiming} target the last-level cache (LLC), which is shared by all cores in a socket.
The evict+reload variant of flush+reload uses cache contention rather than flushing~\cite{liu2015llctiming}.
A less common yet viable strategy corresponds to observing changes in coherence \cite{guru-hpca18} or replacement metadata \cite{kirianskydawg}.
Directories are readily reverse-engineered to construct eviction sets in \cite{mengjia-ieeesp19}.

Recently, multiple security researchers (e.g.,~\cite{Kocher2018spectre,projectzero,Lipp2018meltdown}) have found ways for an attacker to exploit speculative execution to create a transmitter within victim code in order to leak secrets.
Spectre and Meltdown have exploited the fact that code executing speculatively has unrestricted access to any registers and memory, and a host of variants of these attacks have been proposed.
These attacks have motivated \name.

\subsection{Defending against side channel attacks}

Set partitioning, i.e., not allowing occupancy of any cache set by data from different protection domains, can disable cache state-based channels. It has the
advantage of requiring no new hardware, provided groups of sets are allocated at
page granularity \cite{lin2008coloring,zhang09eurosys-coloring} via page
coloring~\cite{taylor1990coloring, kessler1992coloring}.
Sanctum uses set partitioning to block cache timing attacks, as
does \name.

Intel's Cache Allocation Technology (CAT)~\cite{herdrich16hpca-qos,intel-cat}
and its variants (e.g., CATalyst \cite{liu2016catalyst})
provide a mechanism to configure each logical process with a \textit{class of service}, and allocates Last Level Cache (LLC) cache ways to logical processes, falling somewhat short of isolation.
In CAT, access patterns may leak through metadata updates on hitting loads, as the replacement metadata is shared across protection domains.
DAWG \cite{kirianskydawg} endows a set associative structure with a notion of protection domains to provide strong isolation. Unlike CAT, DAWG explicitly isolates cache replacement state. 
Page coloring can be replaced with DAWG or another cache isolation mechanism in \name.


A alternative strategy to protect against cache timing attacks separate from
partitioning is to randomize interference in the cache (e.g.,
RPcache \cite{wang2007bscache, kong2008nobscache}, Random Fill Cache \cite{liu2014randomfill}, Newcache \cite{newcache2016}, CEASER \cite{moin-micro-2018}), CEASER-S \cite{moin-isca-2019},
or alter replacement policies (e.g., SHARP \cite{yan17isca-sharp}, RIC~\cite{kayaalp17dac-ric}, \cite{domnitser2012nomo}).
In the \name design,
we do not allow adaptivity of cache area allocated to an enclave to protect
against cache occupancy attacks (e.g., \cite{cache-occupancy}) and therefore
we cannot use these techniques.


\name uses techniques similar to \cite{ferraiuolo2017full} to achieve timing independence (non-interference) in its memory system. To support demand paging in a secure fashion in \name enclaves, page-level ORAM as briefly described in Sanctum or the more efficient
approach of InvisiPage \cite{invisipage} can be used.

\subsection{Secure Processors}

A detailed review of secure processors is provided in
\cite{costan2016sgxexplained}. Early architectures such
as XOM~\cite{lie2000xom}, Aegis \cite{suh2003aegis}, ARM's TrustZone~\cite{alves2004trustzone}, and Bastion \cite{champagne2010bastion} did not consider
side channel attacks in their threat model.

Intel's SGX~\cite{mckeen2013sgx, anati2013sgx} adapted the ideas in XOM, Aegis and Bastion to multi-core processors with a shared, coherent last-level cache.
SGX does not protect against cache timing attacks, nor control flow
speculation attacks \cite{foreshadow}.
Iso-X extends SGX enclave allocation and does not
protect against cache timing attacks \cite{evtyushkin2014isox}.



Sanctum \cite{costan2016sanctum} introduced a straightforward software-hardware
co-design to provide enclaves resistant to software attacks, including
those that exploit shared
microarchitectural state in a simple in-order processor.
\name is similar to Sanctum, but protects against a broader class of attacks
including those that use the cache/cache directory/DRAM controller bandwidth channels.
The cache hierarchy of \name corresponds to a modern processor as compared to the cache hierarchy modeled by Sanctum, which did not include queues or arbitration logic. Further,
speculation in \name provides a rich attack surface. \name introduces
a microarchitecture-independent
\verb|purge| instruction to flush the out-of-order pipeline.
Keystone \cite{keystone} borrows from Sanctum to
implement enclaves targeting standard RISC-V in-order processors using RISC-V's physical memory protection primitive \cite{waterman2015riscvsystem}.


Komodo \cite{komodo-sosp-2017} builds a privileged
software monitor on top of ARM TrustZone that implements enclaves.
Komodo, as described in \cite{komodo-sosp-2017},
does not support multi-processor execution, and runs on an in-order processor.
InvisiSpec \cite{invisispec} makes speculation invisible in the data cache hierarchy. This comes at a significant performance cost, and does not preclude
speculation based attacks on other shared microarchitectural state.

\section{Baseline OOO Processor}
\label{sec:baseline}

\name is based on the open-source \emph{RiscyOO} speculative OOO processor~\cite{riscyoo-2018}.
RiscyOO implements most features of modern microprocessors, including register renaming, branch prediction, non-blocking caches and TLBs, and superscalar and speculative execution.
In particular, RiscyOO can issue a load speculatively even when older instructions have unresolved branches or memory addresses.
RiscyOO also has a shared L2 cache which is coherent and inclusive with the private L1 caches of each core.
We also refer to the shared L2 as the last level cache (LLC).
Therefore, RiscyOO contains cache-timing side channels and is vulnerable to speculation-based attacks like Spectre.
The detailed configuration of RiscyOO used in this paper can be found in Figure~\ref{fig:riscyoo-config} in Section~\ref{sec:evaluation}.

RiscyOO uses the open-source RISC-V instruction set~\cite{riscv}, and has been prototyped on AWS F1 FPGAs.
Its FPGA implementation boots Linux and completes SPEC CINT2006 benchmarks with the largest input size (\textbf{ref} input) in slightly more than a day.
RiscyOO is a great platform that we can build secure enclaves on: it is vastly more complex than the implementation of Sanctum in \cite{costan2016sanctum}, and presents new challenges for security.

\vspace{-0.05in}
\subsection{Hardware Modifications}\label{sec:hw-changes}

Enclave support in a modern processor system require three interventions:
1. Physical address protection and isolation through the memory hierarchy.
2. A rigorous implementation of a ``purge'' operation to scrub each type of physical resource that can be separately allocated to an enclave.
3. A speculation guard for the security monitor: this software has access to all physical addresses, and must not speculatively load addresses, nor should it speculatively fetch outside of the security monitor's own binary.

Specifically, we summarize the the hardware modifications used by \name{} to support enclaves below.
The details of each modification will be discussed in the rest of paper.

\begin{compactitem}
    \item Flushing microarchitectural states:  Sections~\ref{sec:transition-protect-domains} and~\ref{sec:perf-flush}.
    \item Page-walk check: Section~\ref{sec:steady:addr}.
    \item Turning off speculation and checking instruction fetches in machine mode: Sections~\ref{subsec:security-monitor} and~\ref{sec:perf-nonspec}.
    \item LLC set-partitioning: Sections~\ref{sec:steady:timing} and~\ref{sec:perf-partition}.
    \item MSHR partitioning in LLC: Sections~\ref{sec:steady:timing} and~\ref{sec:perf-mshr}.
    \item Sizing LLC MSHR: Sections~\ref{sec:steady:timing} and~\ref{sec:perf-mshr}.
    \item Other LLC changes to block side-channels: Sections~\ref{sec:llc:solution} and~\ref{sec:perf-arbiter}.
\end{compactitem}

\newpage
\section{Steady State Isolation}
\label{sec:steady-state-isolation}

When examining the interactions between two programs, there are two cases to consider: the first is when the two programs are running on different cores (this section), and the second is when the two programs run on the same core but at different times (cf. Section \ref{sec:transition-protect-domains}).

From an ISA-level point of view, two programs are independent if a program's output does not depend on another program.
We call this \emph{architectural isolation}.
This captures the interaction of instructions in the ISA such as loads and stores, but it does not include side channels such as timing.

Beyond the basic operations provided by the ISA, we assume that the precise time of any microarchitectural event within a core (instruction fetch, issue, execute, commit, etc.) can be measured by the program running on the core.
This conservatively abstracts a cleverly engineered program's ability to measure latencies.



First, we define our guarantee of \emph{weak timing independence} depending on the underlying cause of variation in timing.
If the timing variation is due to one program waiting multiple cycles for another core to release a reserved resource, we define \emph{major timing leak}.
On the other hand, if the timing variation is due to two programs compete for a resource within a single clock cycle, requiring per-cycle arbitration, we define \emph{minor timing leakage}.
Two programs have \emph{weak timing independence} if they are architecturally isolated and only exhibit minor timing leakage (cf. Sections \ref{sec:steady:timing} and \ref{sec:steady:addr}).
Two programs have \emph{strong timing independence} if they are architecturally isolated and have neither flavor of timing leak (cf. Section~\ref{sec:llc}).


In Sections \ref{sec:steady:timing} and \ref{sec:steady:addr}, we ensure that programs using disjoint protection domains have weak timing independence. Section~\ref{sec:llc} achieves strong timing independence.
If two programs use resources from disjoint microarchitectural protection domains, then they are timing independent.

\noindent\textbf{Definitions and Assumptions:}
We consider a \emph{program} to be the collection of all the instructions running on a core in supervisor and lower privilege modes and the corresponding initial data, and we assume no machine mode code is run as part of the enclave program or on the program's behalf during the program's lifetime.
We ignore machine mode because software in machine mode can tamper with arbitrary configuration registers and alter active protection domains.
Including supervisor mode in our analysis simplifies the problem because we do not have to worry about what the operating system is doing to provide services such as virtual memory to user mode programs.
All configuration and usage of virtual memory falls entirely within a program to keep the security monitor as lean as possible and so minor page table operations do not cause an enclave exit to the security monitor.

Each program has a set of physical addresses it accesses.
There are many ways a program can access a physical address.
When virtual memory is off, a physical address is accessed for each instruction fetch and for each load and store.
When virtual memory is on, physical addresses are also accessed for page table walks.
Also since RiscyOO is an aggressive out-of-order execution processor, speculative instruction fetches, speculative loads, and speculative page table walks also cause physical memory accesses even if the speculation was incorrect.
When talking about the set of all physical addresses accessed by a program, we mean the physical addresses of all the above physical memory accesses, even the speculative ones.

We also assume that the entire address space is normal memory, not memory-mapped I/O since we do not trust devices and drivers.

\vspace{-0.1in}
\subsection{Establishing Architectural Isolation}

If two programs do not share any addresses, then without using the timing of microarchitectural events, the execution of one program cannot affect the other.
Therefore, disjoint address spaces imply that programs are architecturally isolated.

\vspace{-0.1in}
\subsection{Establishing Weak Timing Independence}\label{sec:steady:timing}

Unfortunately, having disjoint address spaces between programs, while enough for architectural isolation, is not enough even for weak timing independence.

\noindent\textbf{Cache Partitioning:}
As an example, consider programs $p_1$ and $p_2$ which access disjoint address spaces.
If the two programs access physical addresses $a_1$ and $a_2$ in the same L2 cache set, then accesses from $p_1$ to $a_1$ can cause $a_2$ to get evicted causing $p_2$ to see a miss instead of a hit next time $p_2$ accesses $a_2$.

This issue stems from the two programs dynamically sharing resources (in this case entries in the same cache set).
The on-demand transition of resources from one program to the other is observable by the programs and can therefore be used to infer the demand of the other program.
In order to get around this problem and achieve weak (or strong) timing independence, caches need to be statically partitioned between programs.

MI6 partitions the cache through set partitioning.
Similar to Sanctum~\cite{costan2016sanctum}, \name{} divides equally the physical memory (DRAM) into multiple contiguous \emph{DRAM regions}.
MI6 modifies the LLC cache indexing function so that each pair of DRAM regions map to disjoint cache sets.
That is, the higher bits of the original LLC index are replaced by the DRAM-region ID, which is the highest bits of the physical address (e.g., the highest 6 bits for 64 DRAM regions).
Two programs using set partitioning must only use physical addresses that map to disjoint DRAM regions in order to avoid timing leakage through dynamic sharing of the cache.

\noindent\textbf{MSHR Partitioning:}
The L2 cache in RiscyOO can only handle a fixed number of requests at a time.
These requests are tracked using \emph{miss status handling registers} (MSHRs).
If there are no free MSHRs, then the L2 cannot take any more requests and causes multiple cycles of backpressure for the child L1 caches trying to send requests.

Consider programs $p_1$ and $p_2$ where $p_1$ is causing many cache misses and its requests fill the L2's MSHRs.
If $p_2$ then causes a cache miss, there will be no MSHR available for $p_2$ and the request will be stalled.
This is a timing variation in $p_2$ that is caused by $p_1$ and therefore a major timing leak.

To avoid this timing leak, \name{} partitions the MSHRs in the LLC.
Since only processes that are actively running on the processor cores can occupy MSHRs, \name{} divides equally the MSHRs in the LLC by the number of processor cores, and statically associates each MSHR partition with a processor core.
$p_1$ using all of its allocated MSHRs will not affect whether or not a request from $p_2$ is stalled due to backpressure.

\noindent\textbf{Sizing the MSHRs to Avoid DRAM Backpressure:}
Say that $d_{max}$ is the maximum number of outstanding requests the DRAM controller can handle.
After $d_{max}$ in-flight requests, the DRAM controller asserts backpressure and prevents further requests from being enqueued into it.
If the DRAM controller is asserting backpressure, then DRAM requests will get delayed causing a major timing leak.

In RiscyOO, every request sent to the DRAM controller comes from an L2 request in an MSHR.
Moreover, each request put into an MSHR can send up to two DRAM requests during its lifetime: one for a possible write back and one for a read request.
Therefore, for a machine with a DRAM controller accepting $d_{max}$ outstanding
requests, the number of MSHRs in the cache needs to be
at most $d_{max}/2$. That amounts to about $d_{max}/(2N)$ MSHRs per core, where $N$ is the number of cores.


\noindent\textbf{DRAM Controller Latency:}
Another complication of ensuring timing independence is the DRAM and the DRAM controller.
DRAM controllers often reorder requests so that requests to the same bank are done back-to-back to increase the achieved bandwidth of the DRAM.

Consider programs $p_1$ and $p_2$ where $p_1$ is accessing addresses in DRAM bank 0, and $p_2$ is accessing addresses in DRAM banks 1, 2, and 3.
If $p_1$ and $p_2$ send interleaved requests to DRAM, and $p_1$'s requests were all to the same bank, a reordering DRAM controller would perform $p_1$'s requests back-to-back, changing the timing of $p_2$'s requests.
This changes the timing of $p_2$ based on $p_1$, breaking weak timing independence between the two programs.

For weak (or strong) timing independence, \name must either use a DRAM controller with a constant latency or use a more sophisticated DRAM controller that is aware of protection domains and associated DRAM regions and ensures timing independence across protection domains. That is, optimizations such as row buffer are allowed within a protection domain but not across protection domains.
The DRAM controller model RiscyOO used for evaluation has a constant latency, and we leave the exploration of variable latency DRAM controllers that are timing independent to future work.

\vspace{-0.1in}
\subsection{Address Validation for Protection Domains}
\label{sec:steady:addr}

%

Unlike Sanctum, in \name an enclave does not share virtual address space with untrusted software, as explored in Section~\ref{subsec:security-monitor}.
Coupled with the mechanism of routing page faults to an enclave, and per-enclave page tables, this blocks the page fault and page access
side channel and prevents attacks such as \cite{xu15oakland-tlb,bulck-page-attack},
where the untrusted OS views page faults or accesses.

In order to achieve timing independence, programs need to be able to ensure they only use certain cache sets and no other programs use those sets.
This restriction on address usage includes \emph{all} accesses to memory.
It is easy to write a program that only performs loads and stores on certain addresses, but much harder to ensure that the program will not emit speculative accesses or page table walks to memory that fall outside that range.
It is also much harder to ensure another program will stay outside of your address range.

To make set partitioning easier, \name has hardware support to ensure all physical accesses fall within the specified DRAM regions (and therefore cache sets) allocated to the running program.
Each core in \name has a machine-mode modifiable bitvector containing a bit for each DRAM region determining if that region can be accessed or not.
If the program makes an access (speculative or non-speculative) to an address outside the allocated cache sets, the core will not emit the access to that location and will raise an exception if that access ends up becoming non-speculative.

We need to check the DRAM region for each physical cache access.
\name performs an optimization to simplify the design by caching DRAM region permissions in the TLB.
Each DRAM region is large enough and has proper alignment so that no 4 KB page falls in two DRAM regions.
Therefore, if a page table walk determines an access to a specified page is legal, the translation is added to the TLB and the accesses using the translation are all legal until the DRAM region allocation changes.
To support programs using physical addresses, the security monitor configures cores to trap on virtual memory management instructions so the security monitor can swap in an identity page table when programs try to turn off virtual memory.
Section~\ref{subsec:security-monitor} describes how the TLB is maintained to ensure state transitions do not violate isolation.

\subsection{Achieving Strong Timing Independence}\label{sec:llc}

This section presents the remaining modifications required to achieve strong timing independence, and therefore Property \ref{security-property}.
The mechanisms introduced in Section~\ref{sec:steady:timing} cannot achieve strong timing independence primarily because of the shared LLC.
We first explain the structure of LLC in RiscyOO in Section \ref{sec:llc:struct}, then show the possible minor leakages that break strong timing independence in Section \ref{sec:llc:leak}, and present our solution in Section \ref{sec:llc:solution}.
We also analyze the performance cost qualitatively in Section \ref{sec:llc:perf}.

\subsubsection{Understanding the LLC of RiscyOO}\label{sec:llc:struct}

\begin{wrapfigure}{l}{0.4\columnwidth}
    \centering
    \includegraphics[width=0.4\columnwidth]{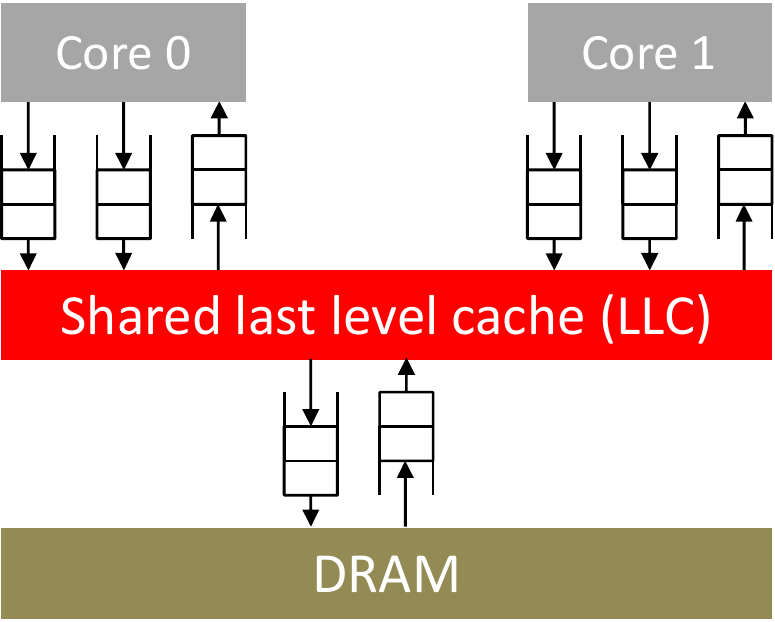}
    \caption{Integration of LLC in RiscyOO}\label{fig:llc-connect}
\end{wrapfigure}

\begin{figure}[!htb]
    \centering
    \includegraphics[width=\columnwidth]{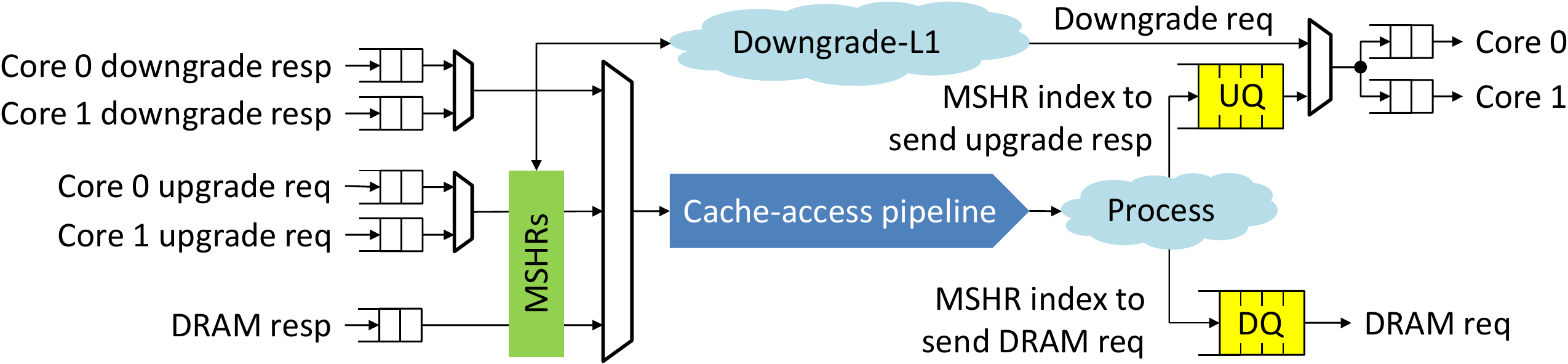}
    \caption{Internal microarchitecture of LLC in RiscyOO}\label{fig:llc-internal}
\end{figure}

Figure~\ref{fig:llc-internal} shows the internal details of the LLC.

Figure~\ref{fig:llc-connect} shows how the LLC is integrated in a two-core RiscyOO machine.
The LLC uses an MSI directory-based cache coherency protocol~\cite{murali2015cache}, and it
uses a dedicated link to communicate cache-coherence messages with the L1s in each processor core.
Each link contains three independent FIFOs to transfer (1) upgrade requests from the L1, (2) downgrade responses from the L1, and (3) upgrade responses and downgrade requests from the LLC, respectively~\cite{murali2015cache}.
The LLC is connected to the DRAM controller using a pair of FIFOs, and the DRAM controller sends responses only for reads.

The LLC contains MSHRs and a cache-access pipeline.
Every incoming message sent to the LLC, including L1 upgrade requests, L1 downgrade responses and DRAM responses, needs to go through the cache-access pipeline to access the tag and data SRAMs of the LLC.
An upgrade request from L1 also needs to reserve an MSHR entry before entering the pipeline.
A DRAM response is buffered in the MSHR entry that initiates the corresponding DRAM read request before it enters the pipeline, and thus \emph{there is no backpressure on DRAM response}.

After a message finishes accessing the SRAMs in the pipeline, it is processed at the end of the pipeline.
After the processing, an L1 upgrade request could be ready to respond, and in this case we enter the MSHR index of the request into a FIFO, i.e., \emph{UQ} in Figure~\ref{fig:llc-internal}, and the response data is buffered in the MSHR entry.
The depth of UQ is equal to the number of MSHRs so it will never backpressure the pipeline.
In other cases where a cache replacement or a cache miss occurs, the L1 upgrade request that causes the replacement or cache miss needs to request DRAM.
In this case, we enter its MSHR index into a FIFO, i.e., \emph{DQ} in Figure~\ref{fig:llc-internal}, and buffer the data in the MSHR entry if writeback is needed.
The depth of DQ is also equal to the number of MSHRs, so it will also never backpressure the pipeline.
That is, \emph{the cache-access pipeline can never be backpressured.}

The final piece in the LLC is the \emph{Downgrade-L1} logic in Figure~\ref{fig:llc-internal}.
Every cycle, the logic looks for an MSHR entry that needs to downgrade any L1s, and sends the downgrade request.

\subsubsection{Minor Timing Leakages in LLC}\label{sec:llc:leak}
Section~\ref{sec:steady:timing} has partitioned the storage elements in the LLC, i.e., cache arrays and MSHRs, to prevent major timing leakages.
However, there are still other shared resources in the LLC that are contended for by messages belonging to different cores and potentially different protection domains.
As long as such contention exists, minor timing leakage is possible.
Here, we enumerate all such contended resources.

\noindent\textbf{Entry port of the cache-access pipeline:}
All the incoming messages are contending on the entry into the cache-access pipeline through a two-level mux as shown in Figure~\ref{fig:llc-internal}.
If two messages from two different cores arrive at the LLC at the same time, then one message will block another for a cycle.
This can lead to minor timing leakage.
Contention between different types of messages from different cores can also form a minor leakage, e.g., a DRAM response for a miss by core 0 and a L1 upgrade request from core 1.

\noindent\textbf{Downgrade-L1 logic:}
All the MSHR entries that need to send downgrade requests are contending on the Downgrade-L1 logic to send downgrade requests.
If the arbitration is not fair, then a large number of MSHR entries of core 0 that need to send downgrade requests can block MSHR entries of core 1 from sending downgrade requests.

\noindent\textbf{UQ and Downgrade requests:}
A head-of-line block in UQ caused by a response to core 0 can stall a later response to core 1 in UQ.
The downgrade requests sent by the Downgrade-L1 logic also contend with the responses in UQ on the outgoing port to processor cores.

\noindent\textbf{DQ:}
If an MSHR entry is entered into DQ because of a cache miss without replacement, then it only needs to send one DRAM read request when it is dequeued from DQ.
This will not block the dequeue port of DQ or lead to leakage.
However, if an MSHR entry enters DQ because of the completion of cache replacement, then it needs to send not only a DRAM writeback request, but also a DRAM read request.
This is because the MSHR entry must have missed in the LLC.
In this case, the dequeue port of DQ will be blocked for one cycle in order to send both requests to DRAM.
This block may delay later requests in DQ, creating minor timing leakage.

We have enumerated all instances of contention in the LLC that can create minor timing leakage.
It should be noted that the DRAM-response port is not a source of leakage, even though responses for cache misses from different cores all go through this port.
This is because the DRAM-response port is never backpressured as explained in Section~\ref{sec:llc:struct}.

\subsubsection{LLC with Strong Timing Independence}\label{sec:llc:solution}
\begin{figure}[!htb]
    \includegraphics[width=1.00\columnwidth]{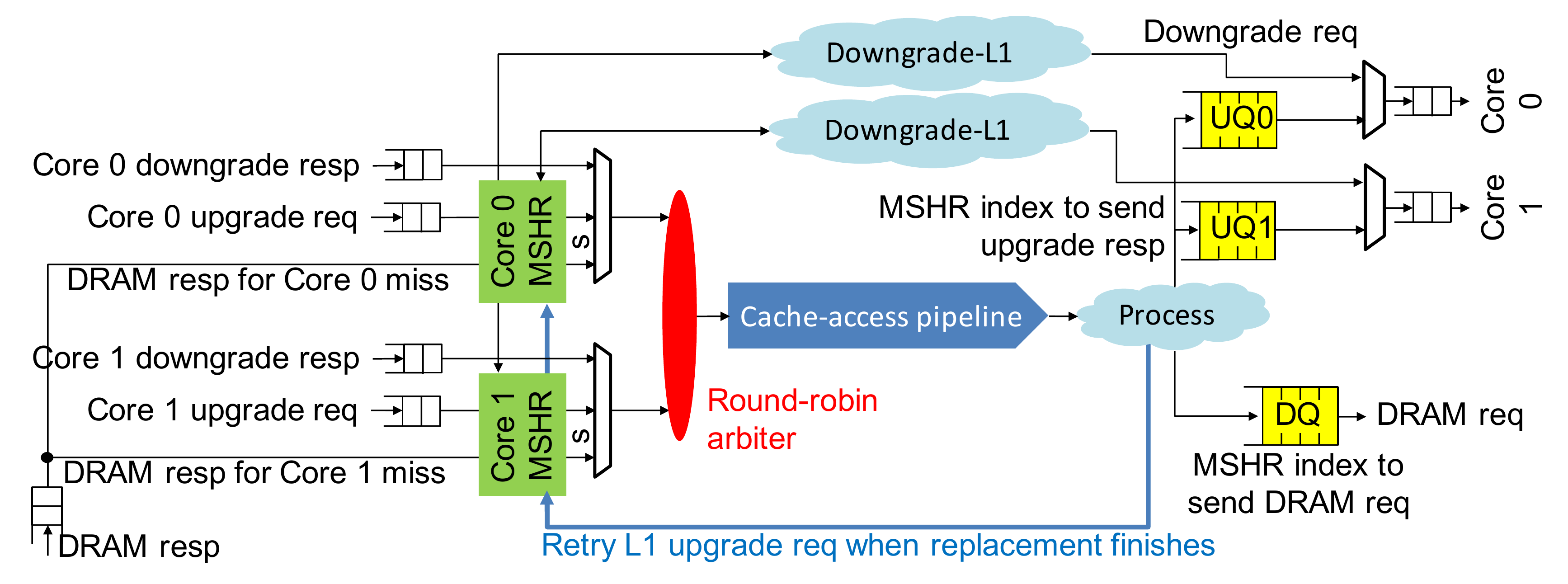}
    \caption{Microarchitecture of LLC in \name{} to achieve microarchitectural  isolation}\label{fig:llc-solve}
\end{figure}

Figure~\ref{fig:llc-solve} shows the microarchitecture of the LLC in \name{} that prevents all the above minor timing leakages and achieves strong timing independence.
We explain the changes made to handle each of the contended resources listed in Section~\ref{sec:llc:leak}.

\noindent\textbf{Entry port of the cache-access pipeline:}
Instead of first merging incoming messages of the same type and then merging different message types, the LLC in Figure~\ref{fig:llc-solve} first merges all incoming messages for the same core, including L1 upgrade requests, L1 downgrade responses, and DRAM responses for the misses caused by the core.
Contention between messages for the same core (and thus the same protection domain) will not cause any leakage.
After merging messages for the same core, we use a \emph{round-robin arbiter} to arbitrate messages from different cores before they enter the cache-access pipeline.
Consider the case where we have $N$ cores with IDs $0\ldots N-1$.
In this case, in cycle $T$, only one message from core $T\%N$ ($T$ modulo $N$) can enter the pipeline.
It should be noted that even if there is no incoming message for this core, messages from other cores cannot proceed.
Since the cache-access pipeline has no backpressure (Section~\ref{sec:llc:struct}), the round-robin arbiter ensures that whether messages from a given core can enter the pipeline is independent from the activity of other cores or protection domains.
That is, strong timing independence is achieved at this entry port.

\noindent\textbf{Downgrade-L1 logic:}
There are two approaches to
solve the contention in the Downgrade-L1 logic.
In the first approach, given
that \name{} has already partitioned the MSHRs across processor cores,
instead of checking all the MSHRs, the Downgrade-L1 logic examines only one MSHR partition for a single core at each cycle.
The logic iterates through all partitions in a round-robin fashion,
providing timing independence.
The logic will still spend a cycle on a partition even if no MSHR in the partition has downgrade request to send.
This ensures that whether the MSHRs for a given core can send downgrade requests is independent from other cores or protection domains.
In the second approach, which we follow, we duplicate the Downgrade-L1 logic for each MSHR partition. Each copy of the Downgrade-L1 logic is therefore only responsible for one MSHR partition, thereby removing any contention (cf. Figure \ref{fig:llc-solve}).

\noindent\textbf{UQ and Downgrade requests:}
To resolve the timing dependence due to UQ, we have split the original UQ into multiple FIFOs (see Figure~\ref{fig:llc-solve}).
That is, UQ$i$ keeps only the MSHR indexes for core $i$, and the depth of UQ$i$ is equal to the size of the MSHR partition for core $i$.
Thus, head-of-line blocking of UQ$i$ is simply a stall within responses for core $i$, i.e., there is no timing dependence across different cores or protection domains.
It should be noted that the total number of entries of all the UQs is still equal to the total number of MSHRs, that is, there is no area overhead.

After the split of UQ, a downgrade request sent to core $i$ can contend only with responses to core $i$ in UQ$i$.
Consider the case that the downgrade request is initiated by an upgrade request from core $j$.
In this case, the address requested by core $j$ is in the same cache set as the downgrading address owned by core $i$.
Thus, cores $i$ and $j$ must be in the same protection domain (see Section~\ref{sec:steady:timing}), e.g., a multithreaded enclave is assigned to cores $i$ and $j$.
Therefore, the contention between UQ and downgrade requests cannot influence timing across protection domains.

\noindent\textbf{DQ:}
The key to solve the problem is to have the dequeue of an MSHR index from DQ always take one cycle, in particular for MSHR entries that are completing cache replacement.
In this way, the dequeue port of DQ will never be blocked, and there is no timing influence with DQ.
Consider an MSHR entry which enters DQ because of completing cache replacement.
In Figure~\ref{fig:llc-solve}, when the MSHR index is entered into DQ, we set a \emph{retry} bit in the MSHR entry.
When the MSHR index is dequeued from DQ, it sends only the writeback request to DRAM, so the dequeue takes only one cycle.
An MSHR entry with the retry bit set will try to re-enter the cache-access pipeline, and enters DQ again as a pure cache miss to issue the DRAM read request.
The re-entry only contends with messages for the same core, and will not create new leakage.
The cache slot is also locked to the MSHR entry so that other upgrade requests cannot occupy the slot.

\subsubsection{Qualitative Analysis of Performance Impact}\label{sec:llc:perf}
First, we note that the split of UQ into multiple FIFOs has zero performance overhead.
The retry of a request that finishes replacement will increase the total processing latency of this request by a few cycles.
However, since the request needs to read DRAM, this increase is negligible.

The Downgrade-L1 logic was duplicated and therefore has
zero performance overhead.
If we had chosen to operate the logic in the round-robin way, then the latency to downgrade L1s may increase proportionately to the total number of cores.

The performance overhead comes mainly from the round-robin arbiter in front of the cache-access pipeline.
The arbiter gives each core $1/N$ of the SRAM bandwidth, where $N$ is the total number of cores.
It should be noted that even without the arbiter, messages from $N$ cores are still contending on the bandwidth of the SRAM.
Therefore, there is no bandwidth loss in the average case.
Performance will decrease if the traffic from each core is bursty.
This is not a big problem for a small multiprocessor with 2 or 4 cores, because the LLC can still take a request from a given core every 2 or 4 cycles, and the core typically will not miss in L1 every cycle.
This issue will be exacerbated as the number of cores increases, but it depends strongly on the timing and contention of memory accesses from different cores.

The arbiter also introduces extra latency in accessing the pipeline, i.e., a message from a given core has to wait for its turn to enter the pipeline.
The average latency is roughly $N/2$ cycles.
We evaluate the performance impact of the increased latency in Section~\ref{sec:perf-arbiter}.

\section{Isolation Across Protection\\ Domain Transitions}
\label{sec:transition-protect-domains}

In addition to achieving isolation of programs scheduled concurrently onto different cores in the system, we must achieve isolation between programs scheduled to use the same core at different times.

As described earlier, \name relies on flushing of microarchitectural state to erase any program-dependent microarchitectural state when scheduling a new protection domain onto the core.
Operationally, we add a microarchitectural \verb|purge| instruction to achieve this; below, we consider the specifics of \verb|purge| by visiting each module it scrubs.
Section~\ref{subsec:security-monitor} discusses how the privileged security monitor, which occupies a dedicated protection domain and executes at highest privilege, orchestrates the low-level operations to implement a secure context switch across protection domains.

\subsection{{\large \texttt{purge}} instruction}
\label{subsec:purge-instruction}

\noindent\textbf{In-flight instructions}:
The out-of-order core consists of many modules containing bookkeeping for in-flight instructions including the register renaming table, register free list, reorder buffer, issue queues, scoreboard, speculation tag manager, load store queue (with corresponding MSHRs), store buffers, and various FIFOs, and smaller modules.
Between contexts, all of these states must correspond to ``no instruction is currently in the processor pipeline'' in order to achieve a comprehensive flush.
The baseline RiscyOO processor correctly flushes these states on privilege change to handle read after write hazards when changing the privilege level.

A multitude of states throughout the modules of the out-of-order core equivalently describe an empty pipeline.
For example, a complete register free list indicates an empty pipeline, but there exist multiple permutations of the free list.
Interestingly, this does not require special consideration so long as these equivalent ``free'' states are not distinguishable by any software.
A similar situation pertains to the issue queue, from where instructions are issued to execute when ready.
In the RiscyOO processor, the issue queue is a circular buffer with associated head and tail pointers.
Any configuration where head and tail pointers are equal maps to an empty state, yet these are entirely indistinguishable by software means.
While not applicable in the case of RiscyOO's issue queue, this module would require additional care to correctly flush program-dependent state in some priority queues, such as the MIPS R10000~\cite{yeager1996mipsr10k}, which favors issuing instructions from low-numbered slots in the queue.

\noindent\textbf{Branch predictor structures:}
Branch predictors were demonstrated as a surface for hijacking speculative execution as part of Spectre \cite{Kocher2018spectre}.
These structures, being deeply stateful, can also transmit information about the control flow of a previously scheduled program via the branch predictions observed after a context switch.
To \verb|purge| the branch predictor state, the branch predictor must reach a well-defined public state, for example via a reset to its initial state.
In order to reduce the overhead of cold branch prediction after each context switch, the processor may opt to implement primitives for saving and restoring predictor state, if practical.

\noindent\textbf{L1 Caches, TLBs, and translation caches:}
Attacks exploiting shared cache tag state to observe secret-dependent changes in memory access latency are a well-explored field, offering increasingly practical examples of private state leakage via the cache.
Any cache timing attack may use not only the shared L2 cache (which \name partitions to address this class of vulnerability), but also the L1 caches, which are time-shared by the programs scheduled onto the same core.
In order to obviate these attack surfaces, L1 caches, along with TLBs (both L1 and L2) and translation caches, must be flushed on context switches.
(TLBs and translation caches are all private to the core.)
In addition to the tag state, the cache lines' replacement policy state must also be scrubbed; RiscyOO fortunately employs a pseudo-random replacement policy with no replacement state.


The TLBs (both L1 and L2) and translation caches use set associative structures with LRU replacement policies.
RiscyOO's implementation of the LRU policy is self-cleaning: when no line's data is present in a set, new lines are filled in a pre-defined order; the act of filling an LRU cache to prime it for eviction scrubs private information in the replacement state.

A noteworthy observation: L2 cache sets need only be scrubbed when re-allocating physical memory. Protection domains use disjoint regions of physical memory, which correspond to disjoint sets in the L2 via set coloring. L2 lines belonging to a de-scheduled protection domain are inaccessible and can remain in the L2 until the domain is scheduled.

\name does not allow for memory shared between arbitrary protection domains; all communication between domains is mediated by a security monitor, as described in Section~\ref{subsec:security-monitor}.

\subsection{Security Monitor Functionality}
\label{subsec:security-monitor}

As in Sanctum~\cite{costan2016sanctum}, we employ a software security monitor to map the high-level semantics of enclaves onto the low-level invariants implemented by the hardware.
While the implementation of a security monitor for \name is not a contribution of this manuscript, this section briefly describes its required functionality, and where it differs slightly from the prior construction~\cite{sanctorum-2018} for Sanctum.
The security monitor interposes on scheduling and physical resource allocation decisions made by the untrusted OS to assert that a given enclave's resources do not overlap with any other software, and to scrub these resources before they are available for re-allocation.
In \name, the security monitor considers two classes of resources: core state, and memory; both include their respective space of subtle side effects these resources have on the memory hierarchy, microarchitecture, and the network-on-chip.
When an OS requests enclaves be scheduled or de-scheduled, the security monitor uses \verb|purge|, as described above, to scrub a core when scheduling enclaves (to create a pristine environment free of adversarial influence) and when de-scheduling enclaves, to erase side effects of enclave execution.
Likewise, before memory is granted to an enclave, or when an enclave is destroyed, the security monitor must be invoked to scrub it before it can be given to a new owner.
During steady-state execution, the security monitor is de-scheduled, and protection domains are isolated via the mechanism is described in Section~\ref{sec:steady:addr}.

The monitor also interposes on an enclave's asynchronous events and exceptions in order to safely de-schedule the enclave before delegating control to the OS handler; the OS observes the event as occurring at the syscall that scheduled the enclave.
The security monitor itself occupies a dedicated protection domain outside the OS's reach, and statically reserves sufficient amount of physical memory for text and data structures implementing the monitor's limited functionality.
The security monitor sets up a physical memory protection primitive (\verb|PAR|: Sanctum's protected address region) to ensure its own integrity from all other software.

All protection domains with the exception of the security monitor execute in virtual memory; the operating system transparently uses an identity page table to access physical addresses.
When protection domains are created or destroyed, stale translations system-wide must be scrubbed: the security monitor forces a TLB shootdown during these transitions to ensure cached translations remain coherent with the system's current security policy.
Because no translations exist to map any virtual address in a protection domain to a physical address outside the protection domain (except as described at the end of the section), speculative fetches and loads will not fall outside protection domain boundaries, and uphold isolation.

As described in Section~\ref{sec:steady:addr}, \name ensures no protection domain may access the memory of another (with the notable exception of the security monitor, which relies on a restricted mode of execution to sidestep any speculative misbehavior, as described below), resulting in straightforward isolation in the L2 cache via page coloring.
Of course, enclaves must occasionally communicate with other software, at a minimum to receive inputs and produce outputs.
While Sanctum and SGX allowed for rich communication between an enclave and its host (the untrusted Linux processor in whose address space the enclaved process exists), \name cannot allow sharing any portion of the address space with untrusted software in order to silo speculative execution.
The security monitor implements explicit messaging between protection domains by allowing a sender to request a message to be copied from the sending domain to a pre-allocated buffer in the receiving domain.
Sanctum's mailbox primitive is one such mechanism, allowing enclaves to send and receive authenticated private 64 Byte messages (local attestation).
\name extends this primitive to also implement a privileged memcopy between an enclave and the untrusted software via an agreed-upon pair of buffers of equal size. The security monitor responds to an enclave's request to ``read'' the OS buffer by copying its contents to the enclave, and to ``write'' it by copying from the enclave's buffer to OS memory.
The security monitor's handling of the primitives above does not depend on the transmitted data, and an enclave's invocation of these APIs is not considered private.
The security monitor therefore need not perform a purge when it mediates communication.
These primitives are a restriction to the more permissive communication mechanism allowed by SGX and Sanctum, so as to defend against timing attacks on shared memory, including those that exploit speculative execution.


Allowing an enclave to interact with the outside world, even only through the security monitor, has implications for our definition of security: instead of leaking its completion time, the enclave also transmits the timing and sequence of its interactions to a potential adversary.
Any communications received from untrusted software are untrusted, and a potential influence channel.
The enclave is responsible for padding the timing of its interactions, and tolerating malicious responses. The padding can be to a constant value for zero leakage, or some value from a fixed size set to limit leakage \cite{fletcher-hpca14}.

Protection domains other than the security monitor (enclaves, untrusted software) share no resources with one another, so side effects of speculative execution within these domains are not visible across protection domain boundaries.
Cores executing these protection domains may therefore speculate with no restrictions.
This is not the case for the security monitor's domain, which executes with highest privilege, and may access arbitrary virtual memory.
As in~\cite{costan2016sanctum}, the security monitor's code is trusted to maintain its own integrity, and to not violate the isolation of other domains.
This trust is insufficient in \name, and we must restrict speculative execution of the security monitor to prevent side effects of mis-speculated fetches and accesses from being observable across protection domain boundaries.
We achieve this by restricting instruction fetch to a range of addresses corresponding to the security monitor, and by throttling register renaming in machine-mode execution (exclusively used by the security monitor), effectively serializing execution.
Restricting instruction fetch prevents the security monitor from leaking information via the shared cache by jumping/branching to a data-dependent address visible to an adversary.
Further, we replicate the security monitor's code within each enclave (these replicas contain nothing confidential and protect their integrity via \verb|PAR|: Sanctum's physical memory protection mechanism), so invoking an enclave's communication primitive does not leave unintended side effects: the monitor's text is not shared by protection domains, and the memcopy is non-speculative (see above), and touches only the two buffers the enclave explicitly intends to access.
A more sophisticated implementation that allows for safe speculation within the security monitor while guaranteeing isolation, and allows finer-grained communication between protection domains, is deferred to future work.

\subsection{Strong Isolation Argument}
\label{subsec:security-argument}

In steady-state execution, the enclave is straightforwardly isolated through its uniquely allocated core and address range.  Since it does not share a core with any other software, its core-local micro-architectural state is private.
Since it does not share an allocated address range with any threads running on other cores, the page invariant circuit ensures external software is not able to access the enclave's physical memory, and the enclave is not able to access addresses outside its address range, including speculative accesses.
Page coloring results in cache set isolation in the shared cache, and MSHR partitioning and other changes result in memory request timing isolation in the cache hierarchy.
The enclaved program is responsible for safeguarding the timing of its public operations.

In transient execution, which includes enclave scheduling, de-scheduling, creation, and destruction, the enclave is isolated by the security monitor sanitizing architectural and micro-architectural state within the core during each event.
When the security monitor is called to perform one of the enclave transient operations, the core switches to machine mode, and that switch causes the pipeline to flush all in-flight speculation from the previously executing program.
The security monitor uses the \verb|purge| instruction to fully flush all the mircoarchitectural state from the processor and it runs a software routine to scrub the architectural state to a known initial state.
Before returning control to software outside machine mode, the security monitor re-purges relevant architectural and micro-architectural state to grant the incoming software a fresh execution environment.

A subtle complication in this process is the security monitor's own unrestricted access to physical memory.
Speculative misbehavior within the security monitor itself might be able to circumvent the isolation of private enclave memory, so we do not speculate in machine mode.

\section{Performance Evaluation}
\label{sec:evaluation}



The performance overheads of \name{} come from (1) flushing per-core microarchitectural states on a context switch, (2) partitioning the shared last level cache (LLC), (3) partitioning and sizing the MSHRs in LLC, (4) the round-robin arbiter for the LLC pipeline, and (5) turning off speculation in machine mode.
Here, we do not evaluate the performance overheads of the changes made in Section~\ref{sec:llc:solution} outside of the arbiter due to the reasons explained in Section~\ref{sec:llc:perf}.
We evaluate each of these five overheads in Sections~\ref{sec:perf-flush} to \ref{sec:perf-nonspec}, respectively, and summarize the overall performance overheads of \name{} in Section~\ref{sec:perf-all}.
Overheads (1) and (5) apply only to enclave applications while overheads (2), (3) and (4) apply to all processes.
Turning off speculation is needed only when the security monitor transfers data on behalf of the enclave to and from the outside world.
We expect these to be rare events which typically only happen at the beginning and end of enclave execution, so we will not take into account the effect of turning off speculation when evaluating the overall performance overheads of \name{} in Section~\ref{sec:perf-all}.

We use the following 7 variants of the RiscyOO processor:
\begin{compactenum}
    \item \emph{BASE}: the baseline insecure RiscyOO processor with parameters listed in Figure~\ref{fig:riscyoo-config}.
    \item \emph{FLUSH}: flushes per-core microarchitectural states at every context switch on top of the BASE processor (used in Section~\ref{sec:perf-flush}).
    \item \emph{PART}: set-partitions the LLC of the BASE processor (used in Section~\ref{sec:perf-partition}).
    \item \emph{MISS}: changes the organization of LLC MSHRs of the BASE processor to model the effect of LLC-MSHR partitioning and sizing (used in Section~\ref{sec:perf-mshr}).
    \item \emph{ARB}: increases the LLC pipeline latency of the BASE processor to model the effect of the round-robin arbiter for the LLC pipeline (used in Section~\ref{sec:perf-arbiter}).
    \item \emph{NONSPEC}: executes memory instructions non-speculatively on top of the BASE processor (used in Section~\ref{sec:perf-nonspec}).
    \item \emph{F+P+M+A}: the combination of FLUSH, PART, MISS and ARB (used in Section~\ref{sec:perf-all}).
\end{compactenum}
We prototyped all 7 processors on AWS F1 FPGAs, and ran SPEC CINT2006 benchmarks with the \textbf{ref} input size.
We did not run benchmark perlbench because we could not cross-compile it to RISC-V.
For benchmarks with multiple ref inputs, we present the aggregate performance number over all the inputs.
In most cases, all benchmarks ran to completion under Linux without sampling.
The only exception is the evaluation of turning off speculation, in which case we truncate the runs because the processor without speculation becomes too slow to finish the benchmarks (see Section~\ref{sec:perf-nonspec}).

\begin{figure}[!htb]
    \centering
    \small
    \begin{tabular}{|l|l|}
        \hline
        Front-end  & 2-wide superscalar fetch/decode/rename \\
                   & 256-entry direct-mapped BTB \\
                   & tournament branch predictor as in Alpha 21264~\cite{kessler1999alpha} \\
                   & 8-entry return address stack \\ \hline
        Execution  & 80-entry ROB with 2-way insert/commit \\
        Engine     & Total 4 pipelines: 2 ALU, 1 MEM, 1 FP/MUL/DIV \\
                   & 16-entry IQ per pipeline \\ \hline
        Ld-St Unit & 24-entry LQ, 14-entry SQ, 4-entry SB (each 64B wide) \\ \hline
        L1 TLBs    & Both I and D are 32-entry, fully associative, \\
                   & D TLB has max 4 requests \\ \hline
        L2 TLB     & Private to each core, 1024-entry, 4-way associative \\
                   & max 2 requests \\
                   & Includes a translation cache which contains 24 fully  \\
                   & associative entries for each intermediate translation step \\ \hline
        L1 Caches  & Both I and D are 32KB, 8-way associative, max 8 requests \\ \hline
        L2 Cache   & 1MB, 16-way, max 16 requests \\
        (LLC)      & coherent with I and D \\ \hline
        Memory     & 2GB, 120-cycle latency, max 24 requests \\
                   & (12.8GB/s for 2GHz clock) \\ \hline
    \end{tabular}
    \caption{Insecure baseline (BASE) configuration}\label{fig:riscyoo-config}
\end{figure}


\subsection{Flushing Per-Core Microarchitectural State}\label{sec:perf-flush}

\noindent\textbf{Methodology:}
We compare the single-core performance of BASE and FLUSH to study the influence of flushing per-core microarchitectural states, including TLBs, L1 caches and branch predictors.
In FLUSH, these states are flushed whenever the processor takes a trap (i.e., an exception or an interrupt) or returns from trap handling.
In one cycle, the hardware can only flush a few entries of L1 caches, TLBs and branch predictors.

Each L1 cache has 512 cache lines, we invalidate one line per cycle.
We cannot invalidate a whole set (i.e., 8 lines) per cycle, because the coherence protocol used in RiscyOO requires L1 to notify L2 even for the invalidation of a clean line.
Entries in TLBs and branch predictors can be discarded directly.
The fully associative L1 TLBs can be flushed in one cycle.
The L2 TLB has 256 sets (each set has 4 entries) and we discard one set per cycle.
For the tournament branch predictor, the largest table has 4096 entries (each of 2 bits), and we discard 8 entries per cycle.
L2 can sustain a bandwidth of one eviction per cycle, though the latency of completing an eviction is larger than one cycle.
All these flushes done in parallel take 512 cycles to complete, during which
time the processor idles.

The stall time for flushing is not the only cost.
When the application resumes from trap handling, the L1 caches, TLBs and branch predictors are all ``cold,'' and it takes time to warm them up.
This will lead to more misses in caches and TLBs, and more branch mispredictions.

\begin{figure}[!htb]
    \centering
    \includegraphics[width=\columnwidth]{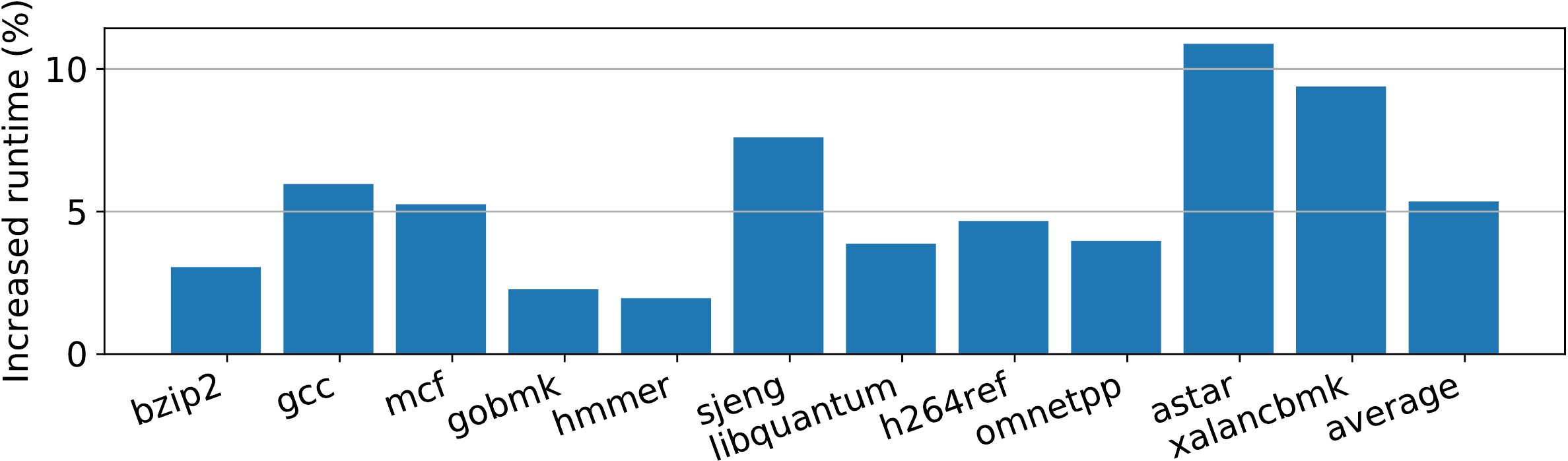}
    \caption{The overall overhead of FLUSH in execution time normalized to the execution time of BASE}\label{fig:flush-cycles}
\vspace{5pt}
    \centering
    \includegraphics[width=\columnwidth]{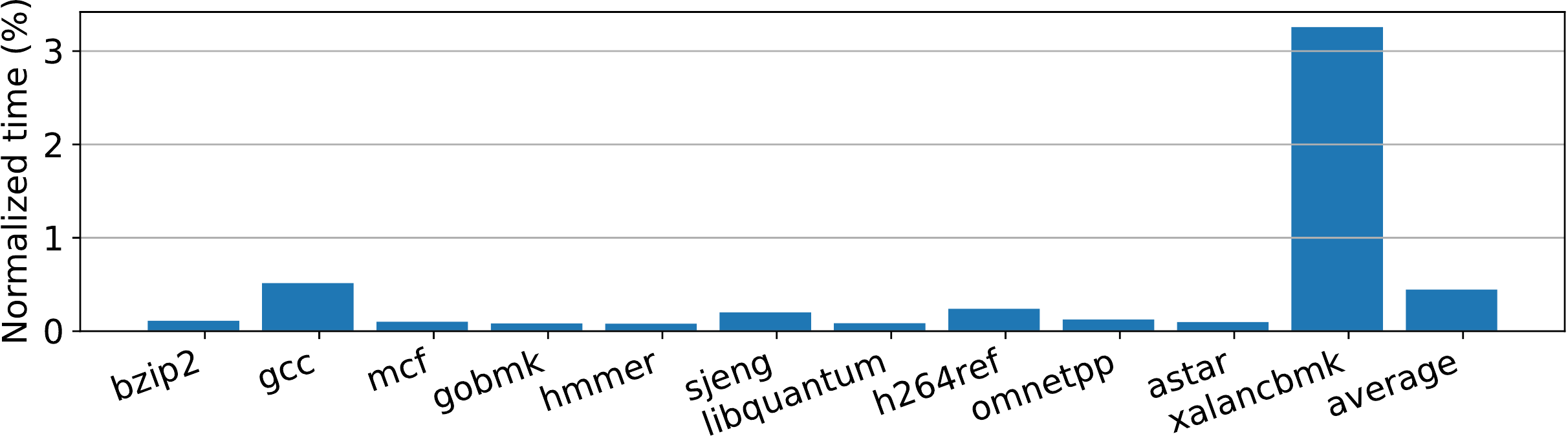}
    \caption{The stall time for flushing states in FLUSH normalized to the execution time of BASE}\label{fig:flush-stall}
\vspace{5pt}
    \centering
    \includegraphics[width=\columnwidth]{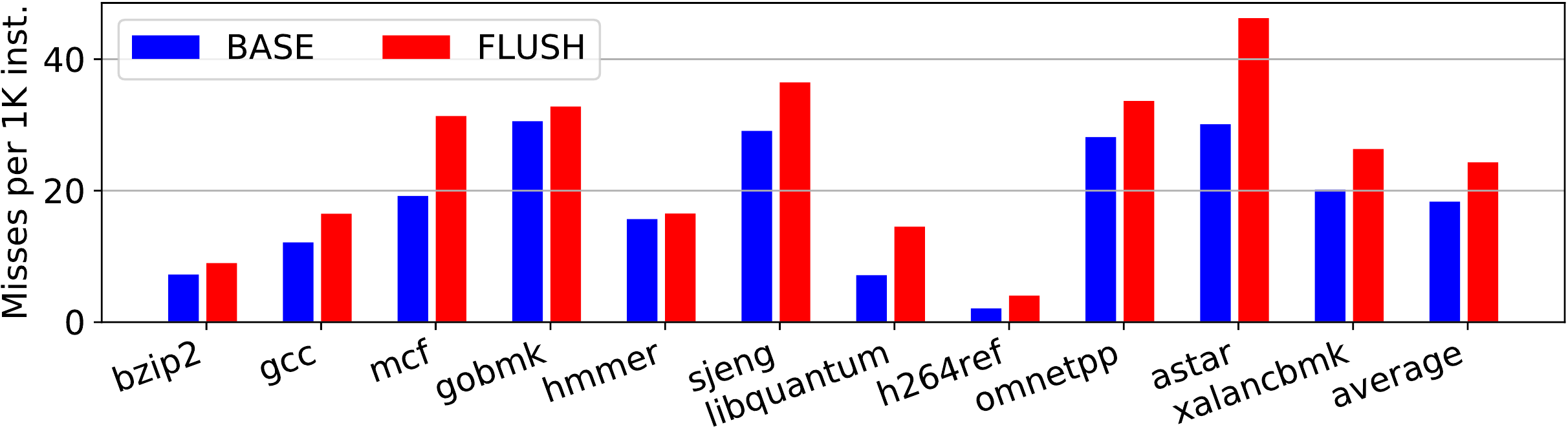}
    \caption{Branch mispredictions in BASE and FLUSH}\label{fig:flush-exe-redirect}
\end{figure}

\noindent\textbf{Results:}
Figure~\ref{fig:flush-cycles} shows the increased execution time caused by flushing for each benchmark.
The last column shows the average across all benchmarks.
Lower bars mean less performance overheads.
The average overhead in execution time is 5.4\%, and the maximum overhead is 10.9\% (in benchmark astar).
As explained earlier, these overheads are caused by (1) the stall time waiting for flushes to be completed, (2) the additional cache and TLB misses caused by the cold start after flushing, and (3) the additional branch mispredictions caused by the cold start after flushing.
We now examine each of these three sources one by one.

Figure~\ref{fig:flush-stall} shows the stall time in FLUSH just waiting for the microarchitectural states to be flushed, and the last column is the average across all benchmarks.
As we can see, flushing the states takes merely 0.4\% of the execution time.
Benchmark xalancbmk has the longest stall time (3.2\%) because it makes a large number of system calls (which trigger exceptions) to print characters to stdout.

While we do not show the results here, the changes in instruction and data cache and TLB misses are negligible, so they are unlikely to be the source of performance degradation.

Figure~\ref{fig:flush-exe-redirect} shows the number of branch mispredictions per thousand instructions in BASE and FLUSH.
The last column is the average across all benchmarks.
It turns that flushing the branch predictions has substantial impact on the misprediction rate.
On average, the mispredictions per thousand instructions rise from 18.3 to 24.3 after microarchitecture-state flushing is enabled.
This significant increase is responsible for the 5.4\% overall performance overhead in Figure~\ref{fig:flush-cycles}.
For benchmark astar, mispredictions go up from 30.1 to 46.2.
This leads to the maximum overhead of 10.9\% in Figure~\ref{fig:flush-cycles}.

\noindent\textbf{Summary:}
Flushing per-core microarchitectural states on context switches has little impact on the miss rates of caches or TLBs, but it increases the branch-misprediction rate substantially.
However, the overall performance overhead caused by flushing is still small.

\subsection{Set-Partitioning the LLC}\label{sec:perf-partition}

\noindent\textbf{Methodology:}
For LLC partitioning, the ideal evaluation methodology would run multiprogrammed workloads on a multiprocessor.
For example, we would like to evaluate a 16-core multiprocessor with 16MB shared L2 cache (LLC).
We assume that each core is still using the parameters in Figure~\ref{fig:riscyoo-config}, and the LLC is still 16-way set-associative and using 64B cache lines.
However, the FPGA does not have enough logic gates and SRAMs for us to prototype such a multiprocessor.
We now explain how to closely approximate the evaluation of a multiprocessor using a single core.

We first point out that the performance overheads of LLC set partitioning mainly come from (1) the decreased LLC size allocated for the enclave application, and (2) the additional conflict misses in LLC caused by changing the cache-indexing function as described below.
In fact, the allocated LLC size is not a concern because our system allows an enclave to claim multiple sets of the LLC.
Besides, running multiprogrammed workloads without considering security would also require partitioning the LLC for Quality of Service (QoS).
This evaluation is not about how to size each LLC partition to achieve the best QoS, so we focus on the second type of overhead, i.e., cache misses caused by changing the indexing function of LLC.

Consider the case that we run a multiprogrammed workload, which consists of 16 SPEC benchmarks, on the 16-core multiprocessor with one benchmark on each core.
For simplicity, we assume the insecure baseline partitions the LLC to allocate 1MB to each core.
If the baseline is using way partitioning, then each core is using effectively a 1MB direct-mapped LLC.
Here, we overestimate the baseline performance by assuming that each core in baseline is using a 1MB cache which still has 16 associative ways. 
That is, the insecure baseline performance can be approximated by the performance of the BASE processor.
In our secure system, consider the case that we set-partition the 16MB LLC into 64 regions, each with 256KB and we assign 4 regions (1MB) to each enclave which runs one SPEC benchmark on a core.
In this case, there is no difference in the allocated LLC size, and the performance overhead of set-partitioning comes only from the change in the cache-indexing function, which we explain next.

Consider a cache-line address $A$ which does not contain the 6-bit line offset (for 64B cache line).
For the baseline insecure 1MB LLC of BASE which has $2^{10}$ sets, the cache index for this line is the lower 10 bits of $A$, i.e., $A[9:0]$.
For the set-partitioned 16MB LLC of our conceptual multiprocessor which has $2^{14}$ sets, the cache index is the combination of the 6-bit DRAM region $R$ and the lower 8 bits of $A$, i.e., $\{R[5:0], A[7:0]\}$.
Since one enclave only gets 4 regions, the higher 4 bits of region $R$ is fixed, and the effective cache index is 10 bits, i.e., $\{R[1:0], A[7:0]\}$.
This is equivalent to indexing the 1MB LLC of BASE using a different index function.

As a result, to approximate the performance impact of LLC set-partitioning, we can simply measure the influence of replacing the higher 2 bits of the LLC index of BASE with the lower 2 bits of the DRAM region, i.e., the effect of changing the LLC index from $A[9:0]$ to $\{R[1:0], A[7:0]\}$.
We refer to the processor that uses the new LLC index as \emph{PART}, and we compare the performance of PART against BASE.

The DRAM region $R$ is the higher bits of the cache-line address.
In the case of 2GB DRAM, $R[1:0]$ would be $A[20:19]$.
The performance overhead of PART is caused by using higher address bits to index the LLC.

\begin{figure}[!htb]
    \centering
    \includegraphics[width=\columnwidth]{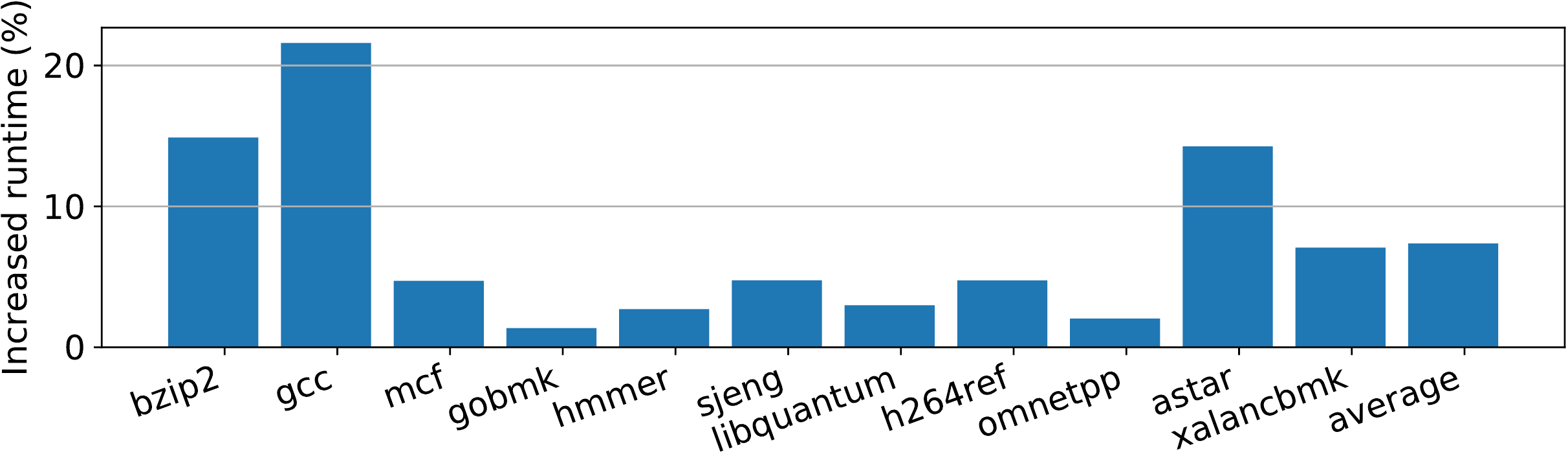}
    \caption{Overhead of PART in execution time normalized to the execution time of BASE}\label{fig:partition-cycles}
\end{figure}

\noindent\textbf{Results:}
Figure~\ref{fig:partition-cycles} shows the increased execution time caused by LLC set-partitioning.
The last column shows the average across all benchmarks.
Lower bars mean less performance overheads.
The average overhead in execution time is 7.4\%, and the maximum overhead is 21.6\% (in benchmark gcc).

\begin{figure}[!htb]
    \centering
    \includegraphics[width=\columnwidth]{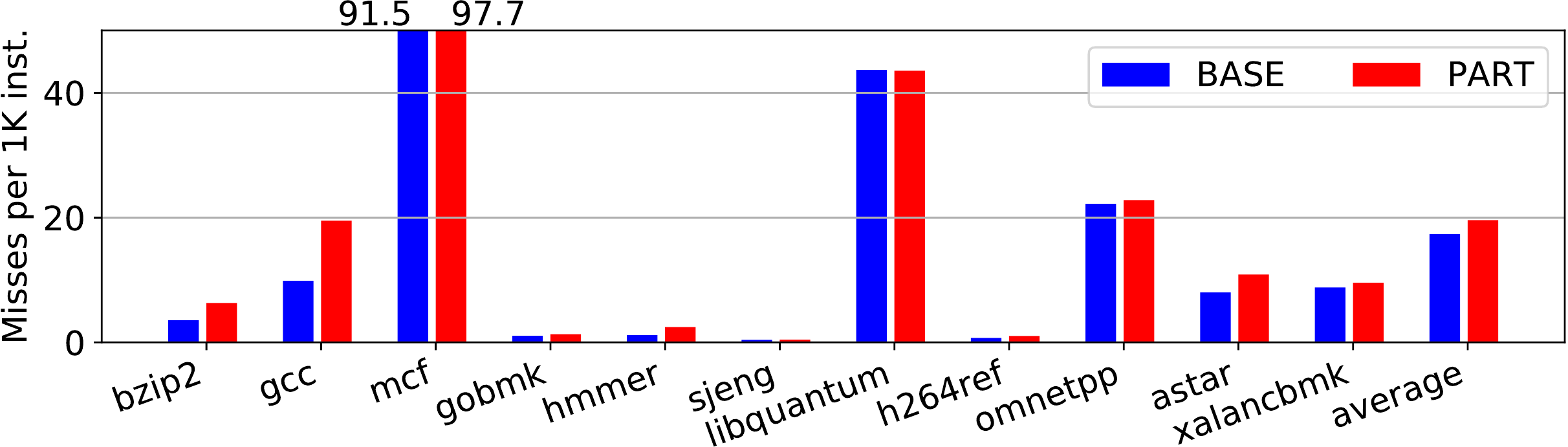}
    \caption{LLC (L2\$) misses in BASE and PART}\label{fig:partition-llc-miss}
\end{figure}

To better understand the performance overhead, we show the number of LLC misses per thousand instructions for BASE and PART in Figure~\ref{fig:partition-llc-miss}.
Because of using higher address bits in the LLC index, the average LLC misses per thousand instructions increase from 17.4 to 19.6 (the last column in the figure).
For benchmark gcc which has the maximum execution-time overhead, its LLC misses get doubled.
These increased misses can be understood as follows.
In our evaluation, we start the benchmark right after the Linux boots.
At that time, most of the physical memory has not been allocated, and Linux tends to allocate physical pages sequentially for the benchmark.
Thus, physical addresses in the working set of the benchmark are likely to share the same higher bits, and thus get mapped to the same LLC index, leading to more LLC conflict misses.

\subsection{Partitioning and Sizing LLC MSHR}\label{sec:perf-mshr}
\noindent\textbf{Methodology:}
We still consider the case of a multiprocessor as in Section~\ref{sec:perf-partition}.
For the insecure baseline, we assume the average memory-system bandwidth available to each core is the same as a single-core BASE processor.
That is, each core in the multiprocessor can occupy 16 LLC MSHR entries and have 24 in-flight DRAM requests on average.

Now consider the secure case that we partition and size the LLC MSHRs to prevent side channels due to contention for LLC MSHRs and DRAM bandwidth.
According to Section~\ref{sec:steady:timing}, the LLC MSHRs should be partitioned to allocate 12 entries for each core, so that the memory requests (including both writebacks and data fetches) generated by each core can never exceed the DRAM bandwidth available to the core (i.e., 24 requests).
If the LLC is organized as several cache banks, then the partition of MSHRs should be done in each LLC bank.
In this evaluation, we consider the case that the LLC is sliced into 4 banks according to the lower bits of the cache index.
In this case, one core will be allocated with 3 entries in each bank (still 12 entries in total).
In the insecure baseline, cache misses by a core can occupy MSHR entries in any bank, i.e., the 16 LLC MSHRs can be distributed across 4 banks in any form.

According to the above analysis, the performance overheads of LLC-MSHR partitioning and sizing come from (1) the reduction in MSHR size, and (2) insufficient MSHRs in a single cache bank (i.e., bank conflicts).
To model the effects of these two overheads, we instantiate the \emph{MISS} processor based on the BASE processor.
The MISS processor has only 12 LLC-MSHR entries compared to 16 in BASE, and the MSHRs in MISS are sliced into 4 banks (according to the lower bits of cache-line addresses).
The performance overheads of LLC-MSHR partitioning can be characterized by the performance difference between BASE and MISS.
The performance of MISS is a pessimistic estimate.
This is because overwhelming one MSHR bank will stall the whole MSHR structure in MISS, while different MSHR banks are independent from each other in the real case.
This modeling error can be avoided by refining the implementation in the future.

\begin{figure}[!htb]
    \centering
    \includegraphics[width=\columnwidth]{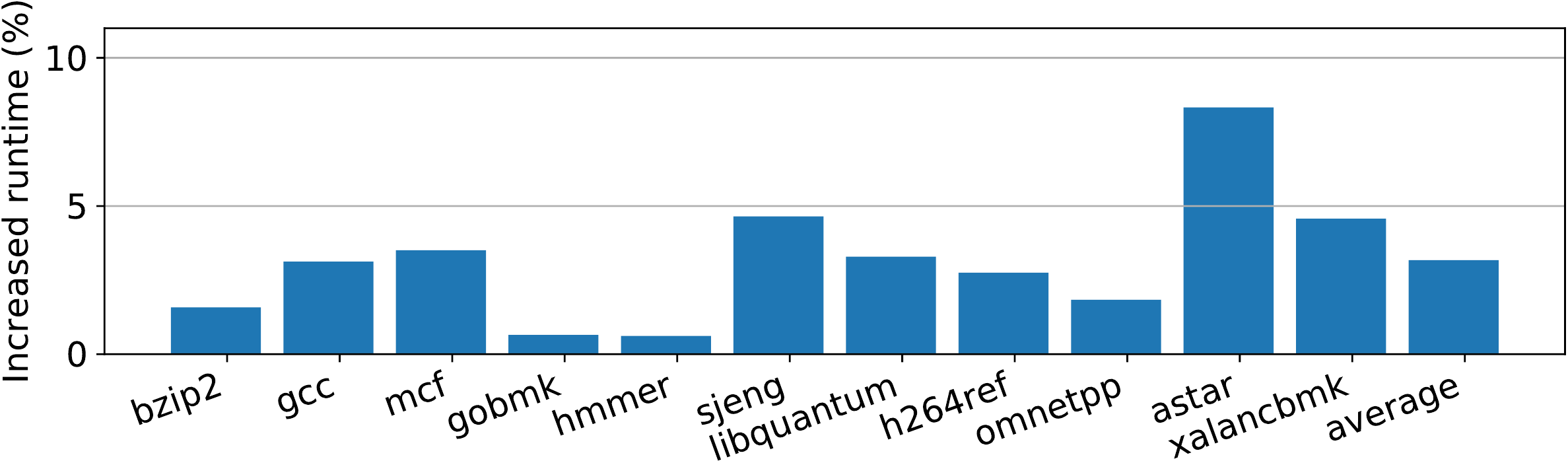}
    \caption{Overhead of MISS in execution time normalized to the execution time of BASE}\label{fig:mshr-cycles}
\end{figure}

\noindent\textbf{Results:}
Figure~\ref{fig:mshr-cycles} shows the increased execution time caused by partitioning the MSHRs in the LLC.
Lower bars mean less performance overheads.
The average overhead in execution time is 3.2\% (the last column), and the maximum is 8.3\% (in benchmark astar).
This overhead is not large.

\subsection{LLC Arbiter}\label{sec:perf-arbiter}
\noindent\textbf{Methodology:}
As described in Section~\ref{sec:llc:perf},
the performance overheads of the round-robin arbiter
associated with the LLC cache-access pipeline
are caused by bandwidth loss in case of bursty cache traffic and the increased latency in accessing the pipeline.
We do not evaluate the overhead due to bursty traffic because it depends strongly on the timing of the concurrently running applications and we are unable to fit a big RiscyOO multiprocessor on an FPGA.
As an approximation, we evaluate only the overhead caused by increased pipeline latency.
For a 16-core multiprocessor, the pipeline latency is increased by 8 cycles on average (cf. Section~\ref{sec:llc:perf}).
We instantiate the \emph{ARB} processor, which increases the LLC-pipeline latency in BASE by 8 cycles, to model the overhead.

\begin{figure}[!htb]
    \centering
    \includegraphics[width=\columnwidth]{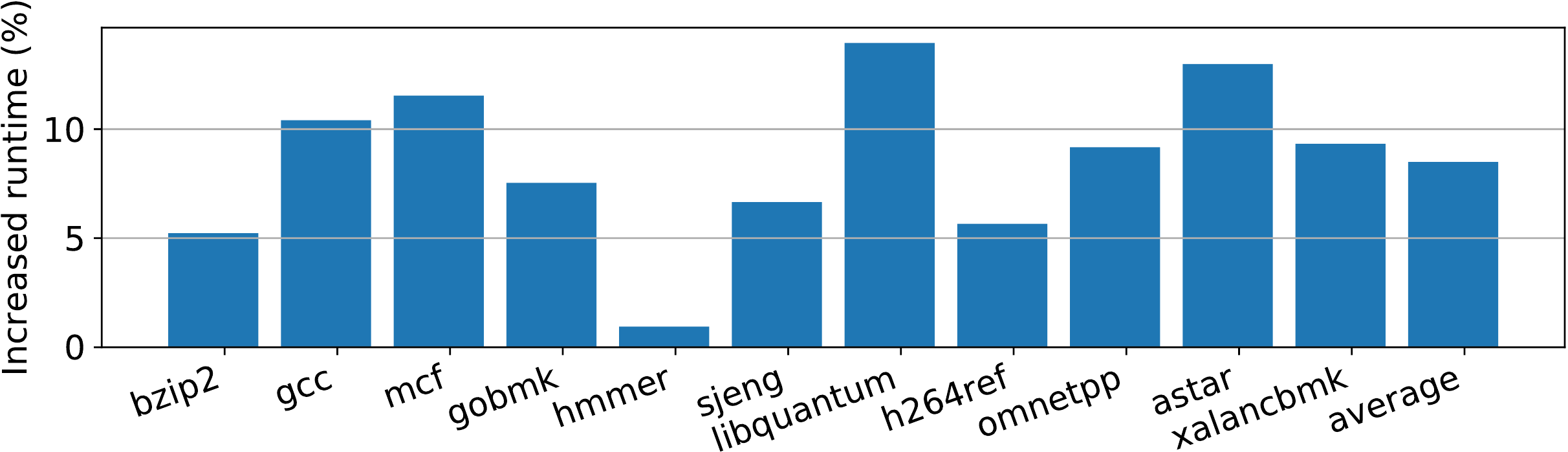}
    \caption{Overhead of ARB in execution time normalized to the execution time of BASE}\label{fig:arbiter-cycles}
\end{figure}

\noindent\textbf{Results:}
Figure~\ref{fig:arbiter-cycles} shows the increased execution time caused by the LLC arbiter.
Lower bars mean less performance overheads.
The average overhead in execution time is 8.5\% (the last column), and the maximum is 14\% (in benchmark libquantum).
This overhead is comparable to LLC set-partitioning.

\subsection{Turning off Speculation}\label{sec:perf-nonspec}
\noindent\textbf{Methodology:}
When a processor runs non-speculatively, the address translation and execution of a memory instruction (e.g., a load or a store) cannot start until the instruction can never be squashed (e.g., by branch mispredictions or exceptions).
Since turning off speculation is uncommon, we implement the non-speculative mode on top of the BASE processor in a simple (but less optimized) way.
In the non-speculative mode, the processor does not rename a memory instruction (and thus cannot enter it into the ROB) until the ROB is empty.
We refer to this processor as \emph{NONSPEC}.

To evaluate the performance overhead of turning off speculation, we run benchmarks on NONSPEC entirely in non-speculative mode.
Since non-speculative mode is much slower than the normal speculative mode, we truncate the benchmarks.
Each benchmark was run for 20 billion instructions without collecting performance data, and then run for 40 billion instructions collecting performance data.
Benchmarks were rerun on BASE using this methodology to get the baseline performance.

\begin{figure}[!htb]
    \centering
    \includegraphics[width=\columnwidth]{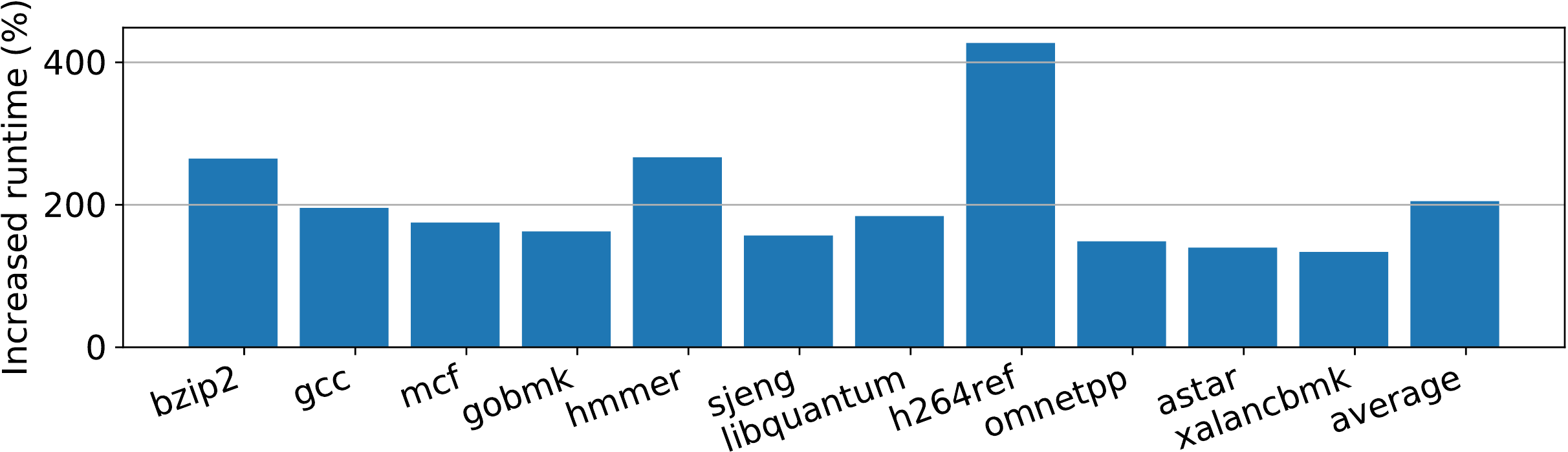}
    \caption{Overhead of NONSPEC in execution time normalized to the execution time of BASE}\label{fig:nonspec-cycles}
\end{figure}

\noindent\textbf{Results:}
Figure~\ref{fig:nonspec-cycles} shows the increased execution time caused by running in the non-speculative mode.
Lower bars mean less performance overheads.
The average overhead in execution time is 205\% (the last column), and the maximum overhead is 427\% (in benchmark h264ref).
Although the overhead is large, it is incurred only at the beginning and end of an enclave program for the common use cases of enclaves, and it does not apply to insecure programs running outside enclaves.

\subsection{Overall Performance Overheads of {\large \bf \name}}\label{sec:perf-all}
\begin{figure}[!htb]
    \centering
    \includegraphics[width=\columnwidth]{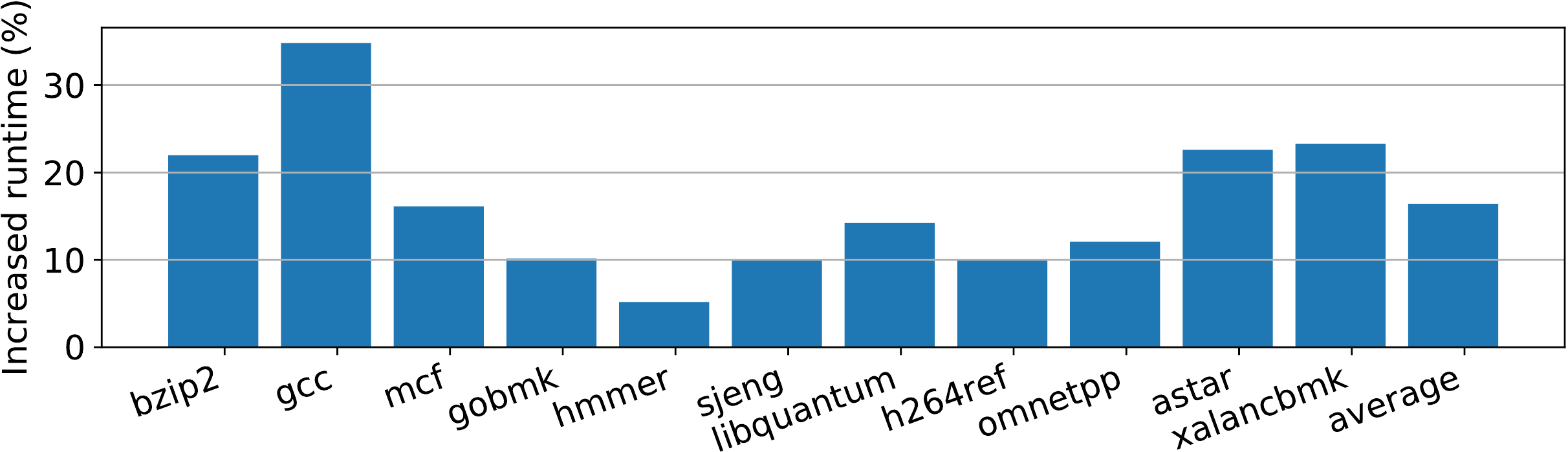}
    \caption{Execution-time overhead of an enclave application in \name{} normalized to the execution time of BASE}\label{fig:secure-cycles}
\end{figure}

\noindent\textbf{Overheads of enclave processes:}
A enclave program running on \name{} is affected by flushing microarchitectural states on every context switch, LLC set-partitioning, LLC-MSHR partitioning and sizing, and the LLC round-robin arbiter.
(We omit the influence of turning off speculation as discussed earlier.)
Its overhead can be approximated by evaluating the performance of the \emph{F+P+M+A} processor, which is simply a combination of FLUSH, PART, MISS and ARB.
Figure~\ref{fig:secure-cycles} shows increased execution time of F+P+M+A compared to BASE, i.e., the performance overhead of an enclave program.
Lower bars mean less performance overheads.
The average overhead of running in the enclave of \name{} is 16.4\% (the last column in the figure), and the maximum overhead is 34.8\% (benchmark gcc).

These performance numbers are a good approximation. The primary omission is ignoring the effect of bursty traffic on the overhead of the LLC arbiter. However, we are conservative in modeling the overhead of LLC-MSHR partitioning and sizing.

\noindent\textbf{Overheads of non-enclave processes:}
Compared to the performance overhead of an enclave program shown in Figure~\ref{fig:secure-cycles}, the overhead of a non-enclave program in \name{} will be less because there is no flushing of microarchitectural states.
\noindent\textbf{Area overhead:}
Our synthesis results show that both BASE and F+P+M+A can be clocked at a maximum of 1GHz.
 Therefore, the additional hardware for enforcing security does not affect clock frequency.
As for area, F+P+M+A is approximately 2\% bigger than BASE.
 Several area-consuming components like LLC, L1 SRAMs and FPUs are not included in the area results, so a 2\% area increase on the rest is quite small.

\section{Conclusion}
\label{sec:conc}

Enclaves strengthen the process abstraction to restore isolation guarantees
under a specified threat model.
Through careful design, prototyping and evaluation, we show how such enclaves
can be supported in \name, an aggressive speculative out-of-order processor
prototype, with reasonable overhead. Further design effort can lower overhead for
unprotected programs by turning on strong timing independence only when at least
one enclave is running, and for unprotected and protected programs
by modifying the OS to reduce the overhead of cache-indexing. A primary 
remaining challenge is allowing enclave software to be more expressive, e.g., 
allowing sharing of memory across protection domains while maintaining isolation.

\section{Acknowledgments}
\label{sec:acknowledgments}
Funding for this research was partially provided by the National Science Foundation under contract number \mbox{CNS-1413920}, Analog Devices, Inc., DARPA \& SPAWAR under contract N66001-15-C-4066 and DARPA under HR001118C0018.

\bibliographystyle{IEEEtranS}
\bibliography{refs,sgx_references,extra_references}

\end{document}